\documentclass[DIV12]{scrartcl}
\usepackage{graphicx}
\usepackage{amsmath}
\usepackage{siunitx}
\usepackage{authblk}
\usepackage{cite}
\usepackage{color}
\usepackage{ulem}
\usepackage[latin1]{inputenc}
\setkomafont{disposition}{\normalcolor\rmfamily}
\setkomafont{caption}{\small\mdseries\selectfont}

\setcapindent{0mm}
\newcommand{\removed}[1] {}

\newcommand{\indur}[2]{#1_{\text{#2}}}
\newcommand{\sbn} {SBN}

\newcommand{\fig}[1]{Fig. \ref{#1}}
\newcommand{\sdist}{\mathcal{A}}
\newcommand{\ldist}{\mathcal{B}}
\newcommand{\FibW}{\mathcal{S}}

\begin{document}
\title{\huge Light localization in optically induced deterministic aperiodic Fibonacci lattices}

\author{\large Martin Boguslawski$^{1,*}$, Nemanja M. Lu\v{c}i\'{c}$^{2}$,  Falko Diebel$^{1},$, Dejan V. Timotijevi\'{c}$^{2}$, Cornelia Denz$^{1}$, and Dragana~M.~Jovi\'{c}~Savi\'c$^{2}$}

\affil{\normalsize
$^1$Institut f\"ur Angewandte Physik and Center for Nonlinear Science (CeNoS), \\Westf\"{a}lische Wilhelms-Universit\"{a}t M\"{u}nster, 48149 M\"{u}nster, Germany\\
$^2$ Institute of Physics, University of Belgrade , P.O. Box 68, 11080 Belgrade, Serbia
\\
$^*$Corresponding author: martin.boguslawski@uni-muenster.de
}
\date{}
\maketitle
\begin{abstract}
As light localization becomes increasingly pronounced in photonic systems with less order, we investigate optically induced two-dimensional Fibonacci structures which are supposed to be amongst the most ordered realizations of deterministic aperiodic patterns.
For the generation of corresponding refractive index structures, we implement a recently developed incremental induction method using nondiffracting Bessel beams as waveguide formation entities.
Even though Fibonacci structures present slightly reduced order, we show that transverse light transport is significantly hampered here in comparison with periodic lattices that account for discrete diffraction.
Our experimental findings are supported by numerical simulations that additionally illustrate a development of transverse light localization for increasing propagation distance.\end{abstract}

\section{Introduction}
Order is one of the central properties to characterize complexly structured systems of any dimensionality.
There is certainly no doubt that periodic systems hold highest order and thus periodicity~-- meaning the invariance in translation for integer multiples of a set of lattice vectors~-- are examined thoroughly for centuries providing nowadays a deep theoretical understanding of waves in periodic media.
Consequently, universal theories such as the Floquet-Bloch theorem were introduced where the model of band structures, for instance, is one prominent achievement recovered in numerous disciplines. 
One of the best established model transfers from electron-wave propagation in solids is covered by the Floquet-Bloch theorem to the field of photonic lattices with light (or electro-magnetic waves in general) as the propagating wave in a periodically modulated refractive index material \cite{Russell, Joannopoulos}.
The existence of band gaps where propagation is forbidden as, per definition, no eigenstates exist within these gaps is closely connected to a band-structure system.

Quasi-periodic structures obviously are of less order than periodic structures as they lack in short-range order \cite{Shechtman, Levine}.
However, the spectra of both, periodic and quasi-periodic structures only present discrete contributions and are distinguishable by their rotation symmetry at the utmost which is limited to 2-, 3-, 4-, or 6-fold for periodic lattices. 
For quasi-periodic structures even higher rotation symmetry can emerge and, more interestingly, band-structure properties can be assigned to these systems as well, de facto offering complete band gaps \cite{Chan, Florescu}.

Numerous experimental techniques to achieve refractive index modulations such as direct laser writing \cite{Pertsch} and photo lithography techniques \cite{Campbell} were suggested during the last decades \cite{Joannopoulos}.
In photorefractive media, for instance, the refractive index is usually modulated by illumination with structured light \cite{Trompeter}.
In particular, optical induction of elongated two-dimensional (2d) photonic structures can be achieved using so-called nondiffracting beams \cite{Durnin} where the intensity is modulated transversely while being constant in the direction of propagation \cite{Rose}.
This technique is highly dynamic as the refractive index modulation achieved with low to moderate intensities is reversible and additionally introduces a nonlinear response allowing for the realization of discrete soliton formations in periodic photonic structures \cite{Christodoulides, Lederer}.
Corresponding to the induction of periodic photonic lattices, the use of quasi-periodic nondiffracting writing beams such as a 5-fold Penrose intensity-configuration allows for the optical generation of according photonic quasi-crystals \cite{Levi}.

In general, nondiffracting beams cover an enormous variety of intensity modulations, ranging with decreasing order from periodic \cite{Boguslawski_dNdB} to quasi-periodic \cite{Rose} to random structures \cite{Boguslawski_AL, Brake}.
Yet, discrete structures without rotation symmetry are barely feasible with a single-beam induction configuration.
We thus presented recently that optical induction techniques can be extended to resemble any aperiodic structure by sets of zero-order Bessel beams as waveguide formation entities \cite{Diebel1}.
Applying this incremental induction technique \cite{Boguslawski_SLatt, Boguslawski_LStrc}, we realized full aperiodic so-called Vogel lattices that are prominent examples for structures inspired by nature as similar arrangements can be found by plant growth following golden-angle phyllotaxis \cite{Trevino}.
Vogel spirals do not show any periodicity in most cases, though they are created deterministically as a construction rule assembles the structure iteratively up to an arbitrary quantity of elements.

Of course, there are uncountable examples of deterministic aperiodic structures \cite{DalNegro3} holding different degrees of order \cite{Baake}.
Yet, in terms of order the Lebesgue's decomposition theorem was suggested to categorize structures along their spectral appearance \cite{Hiramoto, Macia}.
Spectra of aperiodic structures with highest order including Fibonacci patterns \cite{Gumbs, Lifshitz} show only discrete contributions while low-order structures such as Rudin-Shapiro patterns feature continuous quasi-white-noise spectra \cite{DalNegro3}.
Accordingly, structures with intermediate order have both, singular and continuous spectral properties.
The most famous structure among them certainly arises from the Thue-Morse sequence \cite{Baake}.

The question of how an altering degree of order influences wave (or light) propagation remains untouched within this categorization scheme.
As localization in the sense of reduced light transport becomes more significant in disordered photonic systems \cite{Schwartz, Lahini, Jovic1}, the occurrence of localized modes due to a deterministic aperiodicity is ascribed to flat bands along with band gaps that unclose with decreased order \cite{Chan}.
In consequence, finding signatures of localization already in singular-spectral structures would directly refer to an expected reduced degree of order in comparison to periodic lattices with likewise singular spectra.

In this contribution, we examine waveguide patterns of pure-point spectra by exploring light localization effects \cite{Renner, Lucic} in optically induced refractive index Fibonacci structures.
We benefit from a highly dynamic and reversible induction scheme at realizing largely elongated photonic structures that offer excellent interaction distances.
In Fibonacci arrangements, diverse local configurations of lattice sites exist \cite{Lifshitz}.
Hence, we consider an output average for waveguide excitation at different input positions.
By comparing our results with a periodic photonic lattice configuration where discrete diffraction with a high rate of transport is expected to be found \cite{Pertsch}, we underline that order is significantly diminished already in aperiodic structures with pure-point spectra, disabling light transport and causing power-law weighted localization \cite{Gellermann}.

The manuscript is structured as follows.
First, we present our approach of applying one-dimensional Fibonacci words to create 2d lattices with Fibonacci tiling.
After we have described the applied experimental techniques and specific parameters that we used for the optical induction, we present our results of single-waveguide excitation and compare these experimental data with numerical simulations. 
Finally, we cross-check our results against light propagation events in a regular lattice before a conclusion is drawn in the last section.

\section{Design and induction of 2d Fibonacci lattices} \label{sec:expTchnq} \label{sec:FibLattice}
\begin{figure}[h]
  \center
	\includegraphics[width=.50\textwidth]{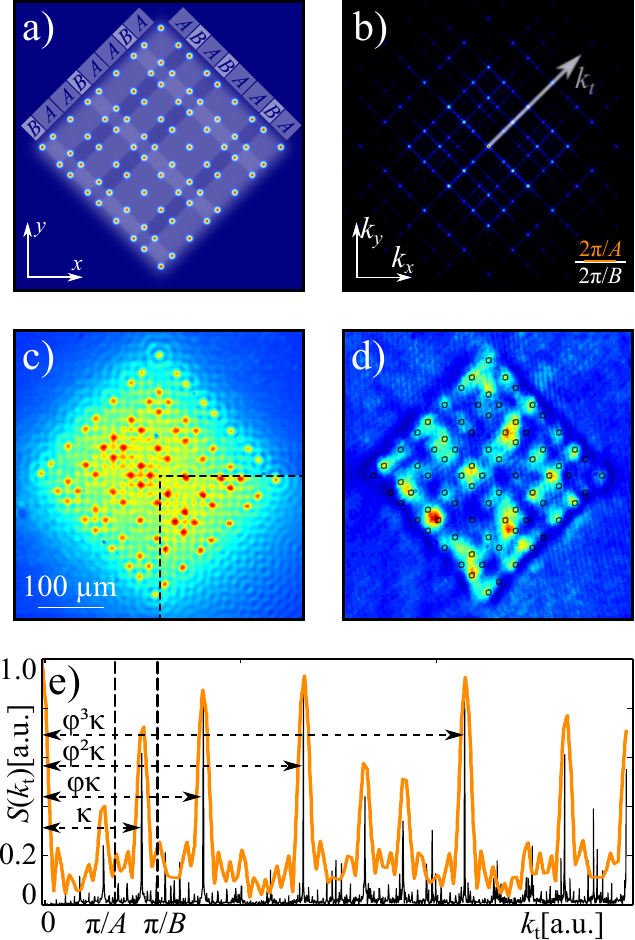}
	\caption{(a) Fibonacci lattice with Gaussian beam sites (underlying Fibonacci words indicated alongside). (b) Spatial spectrum $S(\indur{k}{x}, \indur{k}{y})$ according to lattice shown in (a). (c) Measured effective intensity with Bessel beam lattice sites taken by multiple-shot illumination at back face of crystal. Bottom right quadrant shows according numerical simulation. (d) Experimental output for plane-wave probing (contours indicate waveguide positions). (e) Plot of (orange) spectrum $S(\indur{k}{t})$ along direction denoted in (b) and of (black) ideal spectrum for extended aperiodic Fibonacci lattice with $\delta$-function lattice sites.
	}
	\label{fig:wholeFibLattice}
\end{figure}

Our approach to design an aperiodic Fibonacci pattern is to vary the distances of adjacent wave\-guides encoded as Fibonacci words \cite{Lifshitz}.
These sequences are binary, and according to the Fibonacci series the $n^\text{th}$ word is generated by combining the $(n-1)^\text{st}$ and the $(n-2)^\text{nd}$ word, such as $\FibW{}_n = \{\FibW{}_{n-1} \FibW{}_{n-2}\}.$
Giving the first two words determines the complete set of all words.
We define two different distances $\sdist{}$ and $\ldist{} =  \sdist{}/\varphi$ where $\varphi = (1 + \sqrt{5})/2$ is the golden ratio, and set $\FibW{}_0 = \sdist{}$ and $\FibW{}_1 = \sdist{}\ldist{}$ such that the first five Fibonacci words read as
\begin{eqnarray*}
\FibW{}_0 = \sdist{}, \quad
\FibW{}_1 = \sdist{}\ldist{},\quad
\FibW{}_2 = \sdist{}\ldist{}\sdist{},\\
\FibW{}_3 = \sdist{}\ldist{}\sdist{}\sdist{}\ldist{},\quad
\FibW{}_4 = \sdist{}\ldist{}\sdist{}\sdist{}\ldist{}\sdist{}\ldist{}\sdist{}.
\label{eq:firstFibWords}
\end{eqnarray*}
Picking two sub-words of length $N$ from a very long Fibonacci word $\FibW{}_n$ with $n \gg N$ starting at arbitrary but different elements for two transverse directions, we receive a deterministic aperiodic, non-symmetric structure with $N\times N$ sites as depicted in \fig{fig:wholeFibLattice}(a).
Here, the typical character of a Fibonacci word is present in both orthogonal directions.
That is, distance $\sdist{}$ occurs with a probability of $0.62$, thus more frequently than $\ldist{}$ which yields typical structure groups: quad, double and single waveguide elements.

The Fourier transform of the Fibonacci structure as shown in \fig{fig:wholeFibLattice}(a) gives the spatial spectrum whose absolute value distribution $S(\indur{k}{x}, \indur{k}{y})$ is presented in \fig{fig:wholeFibLattice}(b).
The singular character becomes apparent here but especially in \fig{fig:wholeFibLattice}(e) where the orange plot represents the cross section distribution $S(\indur{k}{t})$ along the arrow drawn in in \fig{fig:wholeFibLattice}(b).
For comparison, the corresponding plot for a largely extended Fibonacci grating and thus with extensive resolution is given by the black curve.
Notice that particular frequency peaks have mutual distance relations equal to the golden ratio $\varphi{}$.
Frequencies with distinct amplitude peaks can be found at $k = \varphi^m \kappa$ where $\kappa$ is the frequency of the prominent peak settled between $\pi/\sdist{}$ and $\pi/\ldist{}$ and $m$ is a natural number.
Thus, in addition to the nomenclature aperiodic, a classification of the presented spectrum as singular or pure-point is appropriate, as well \cite{Macia}.

After we have determined the waveguide positioning scheme resembling a 2d Fibonacci arrangement, we apply the incoherent-Bessel beam induction method to preserve the nondiffracting character of the effective intensity during the writing process.
This method implies to write every single waveguide with an appropriate nondiffracting Bessel beam of zeroth order and structural size $g$.
Determining the structural size $g$ of the Bessel beam intensity settles the waveguide diameter which scales proportionally with $g$.
In our considerations, we fix $g$ to $\SI{13}{\micro\meter}$ and the effective waveguide distance to {$\indur{d}{eff} = \SI{32}{\micro\meter}$}.
This yields $\sdist{} = \SI{37.5}{\micro\meter}$ and $\ldist{} = \SI{23.2}{\micro\meter}$.
A simulation of the resulting transverse effective intensity distribution is given in the lower right quadrant of \fig{fig:wholeFibLattice}(c).

\begin{figure}[t]
  \center
	\includegraphics[width=.70\textwidth]{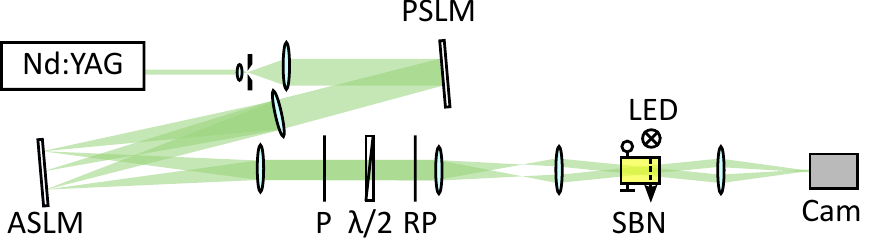}
	\caption{
	Experimental setup for induction of Fibonacci lattices with incoherent Bessel beams. {(A/P)SLM: (amplitude/phase) spatial light modulator, Cam: CCD camera, $\lambda/2$: half-wave plate, LED: background illumination, (R)P: (rotatable) polarizer, SBN: strontium barium niobate crystal; lenses and pin hole not labeled}.
	}
	\label{fig:expSetup}
\end{figure}

In general, our experimental setup incorporating a set of spatial light modulators (SLM) is appropriate to generate any kind of nondiffracting beam \cite{Rose}.
The respective setup scheme is presented in \fig{fig:expSetup} and the induction process corresponds to descriptions given in Ref. \cite{Diebel1}. 
We use SLMs to experimentally realize numerically calculated light fields in a particular image plane defined by an optical imaging system with a demagnification factor of roughly $1/6$.
Particularly, the PSLM (Holoeye \textit{Pluto}) is positioned in real space (related to the image plane) in order to modulate incoming plane waves.
The entire field information of the desired beam is encoded in elaborate diffraction gratings displayed by the PSLM.
An ASLM (Holoeye \textit{LC-R 2500}) is placed in Fourier space for spectral low-pass filtering reasons.
For all experiments, we use a frequency doubled Nd:YAG continuous-wave laser source at \SI{532}{\nano\meter} wavelength.

To resemble the desired $9 \times 9$ Fibonacci structure as presented in \fig{fig:wholeFibLattice}(a), we basically change the position of each Bessel writing beam and put its central maximum to the defined position resulting from the given structure design.
A full set of diffraction gratings containing the necessary writing light fields is generated by convolution of a sparse matrix (with the entire position information) and a basic Bessel beam field distribution.

For an actual experimental induction it is sufficient to send every Bessel beam diffraction grating sequentially to the SLM each generating a corresponding light intensity in the volume of interest where a photorefractive strontium barium niobate (\sbn{}) crystal is placed.
During the illumination, an external field of \SI{2}{\kilo\volt\per\centi\meter} is applied to the crystal \cite{Trompeter}.
This sequential induction scheme implies an effectively incoherent superposition of all writing light fields of one set.
The obvious motivation is that a coherent overlap of contributing Bessel beams would cause undesired intensity modulations since off-diagonal terms of the resulting field would be nonzero.
However, by introducing an effective intensity $\indur{I}{eff}$, we aim to implement the sum of all intensities $\indur{I}{eff} = \indur{\sum}{k} \indur{I}{k} = \indur{\sum}{k} \left|\indur{E}{k}\right|^2$ rather than the absolute square of all fields $\left|\indur{\sum}{k}\indur{E}{k}\right|^2$.

Simulated and experimental effective intensities are given in \fig{fig:wholeFibLattice}(c).
The intensity distributions of writing and probing light fields are recorded by a CCD camera imaging system.
To receive the experimental picture, the output intensities of each writing beam are taken individually and added up afterward.
This overall intensity pattern displays the effective intensity that optically induces the 2d photonic structure.
A LED is placed above the crystal to actively erase the inscribed structure for further light potential inductions.

Figure \ref{fig:wholeFibLattice}(d) shows the output intensity when probing the induced structure with a plane wave.
Notice, that waveguide positions of quad and double elements cannot be resolved properly here as one intensity envelope covers clusters of adjacent waveguides for predominantly perpendicular probe beam incidence.
However, we will see later on that single waveguide excitation indicates accurate induction of waveguide clusters according to the effective intensity distribution.

\section{Light propagation characteristics in Fibonacci lattice} \label{sec:PropInFibLatt}
After the induction, we experimentally investigate the influence of the Fibonacci lattice with its diverse local conditions on the beam propagation in the linear regime of low probing beam power of several $\si{\micro\watt}$, and compare our experimental results with numerical simulations.
In order to numerically model light propagation along the $z$ direction of any photonic lattice, we consider the paraxial wave equation $\left\{2 \i k\partial_{z} + \Delta_{\bot} - k \indur{n^2}{e} r_{33} \left(\partial_{x} \indur{\Phi}{sc}\right) \right\}A(\vec r) = 0$ for a slowly varying electric field amplitude $A(\vec r)$ with wave number $k$.
In this equation, $\partial_{x_i}$ is the spatial derivation in $x_i$ direction, $\Delta_{\bot}$ the transverse Laplacian perpendicular to $z$, $\indur{n}{e}$ the unmodulated refractive index for extraordinary polarization, $r_{33}$ the corresponding electro-optical coefficient, and $\indur{\Phi}{sc}$ denotes a light potential caused by the optically induced internal electric field that in our case holds the Fibonacci structure.
In order to simulate light propagation, we choose parameters that match  experimental conditions and use a split-step method to evaluate the wave equation.

\begin{figure*}[t]
  \center
	\includegraphics[width=.85\textwidth]{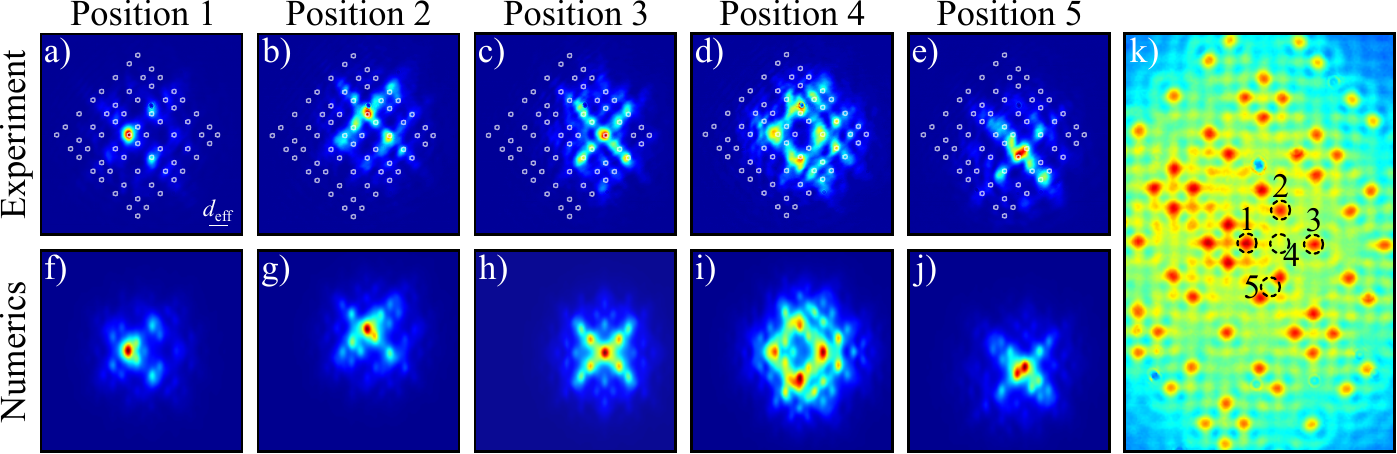}
	\caption{Light propagation in {aperiodic} Fibonacci photonic lattices. Intensity distributions at exit face of crystal experimentally observed (first row) and numerically calculated (second row) for input probe beam size $w_0 = \SI{14}{\micro\meter}$. Columns correspond to input beam positions 1 {to} 5{, as shown in (k)}.
	}
	\label{fig:probing20mm}	
\end{figure*}

For beam propagation studies, a Gaussian probe beam of $w_0 = \SI{14}{\micro\meter}$ beam waist is launched at different input positions of a Fibonacci photonic lattice. 
Figure \ref{fig:probing20mm} shows five selected output results of light propagation through a lattice of $\SI{20}{\milli\meter}$ propagation length.
These different input positions on and between lattice sites are marked from number 1 to 5 in \fig{fig:probing20mm}(k) and relate to output distributions shown in the first to fifth column in \fig{fig:probing20mm} as indicated.
Top-row images present typical output distributions experimentally observed at the exit face of the crystal [\fig{fig:probing20mm}(a--e)] while the bottom row represents the corresponding distributions obtained numerically [\fig{fig:probing20mm}(f--h)].

We notice a very good agreement between experimentally obtained and numerically simulated results.
Moreover, a pronounced heterogeneity of output profiles for different launching positions can be found indicating that the appearance of excited modes is highly diverse.
Naturally, the separation between incident and neighboring lattice sites has a very strong influence on the propagation process as the coupling coefficients vary inversely to the distance \cite{Pertsch}.
Thereby, light is subject to be guided along waveguide lattice sites accompanied by coupling between adjacent waveguides, as well as to ballistic propagation for predominantly high spatial frequencies encountering low spectral amplitudes [cf. \fig{fig:wholeFibLattice}(e)].

\section{Comparison with beam propagation in periodic {lattices}}\label{sec:CompareFibRegLattice}
\begin{figure}[b]
  \center
	\includegraphics[width=.70\textwidth]{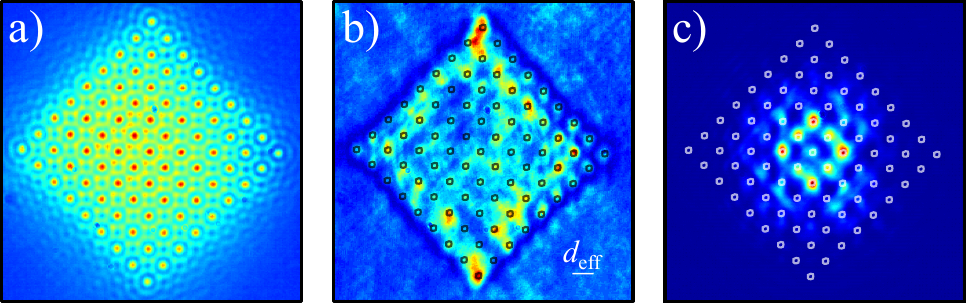}
	\caption{Experimental images for periodic photonic square lattice. (a)~Effective intensity, (b)~plane-wave, and (c)~single-site probing output.
	}
	\label{fig:exp_regular}	
\end{figure}

To substantiate the localization character of Fibonacci structures, we compare this kind of deterministic aperiodic lattice with the case of periodic lattice, both of $\SI{20}{\milli\meter}$ length.
Last named lattices are arranged periodically with as many lattice sites as for the aperiodic case, using again the sequential Bessel beam writing technique to preserve comparability.
For this geometry, the lattice period corresponds to the effective distance of the Fibonacci lattice, {$d=\SI{32}{\micro\meter}$}.
The experimental effective intensity, the plane-wave probing output, and the focused Gaussian beam output are shown in \fig{fig:exp_regular}(a--c), respectively.
Again, probing the structure with a plane wave does not resolve the light potential in very detail as can be found in \fig{fig:exp_regular}(b), though, the image indicates qualitatively the area of modulated refractive index.
A Gaussian beam of $w_0=\SI{14}{\micro\meter}$ coupled into the central lattice point reveals the typical discrete diffraction at the output face as expected in regular lattices \cite{Pertsch}.

We further numerically study the beam propagation in Fibonacci and regular photonic lattices along the direction of propagation ($z \in[\SI{0}{\milli\meter},\SI{40}{\milli\meter}]$) with respect to the development of the probe beam width, having the experimental case of $z = \SI{20}{\milli\meter}$ included.
In such kind of deterministic aperiodic as well as in disordered systems, it is common to perform analysis by averaging over different input beam positions [cf. \fig{fig:probing20mm}(k)] in order to diminish effects of the local environment given by the surrounding conditions \cite{Lucic, Schwartz, Jovic1}. 
For quantitative analysis, we use a relevant measure to characterize the degree of light confinement, the effective beam width $\indur{\omega}{eff}(z)=P(z)^{-1/2}$, where  
\begin{eqnarray*}
P(z)=\frac{\int |E(x,y,z)|^4dx dy}{\left(\int |E(x,y,z)|^{2}dx dy \right) ^{2}}
\label{eq:IPR}
\end{eqnarray*}
is the inverse participation ratio \cite{Schwartz}.
To extract an averaged effective beam width $\indur{\omega}{eff}$ for the Fibonacci structure, we took the arithmetic mean of beam widths for 36 different on-site input positions.

\begin{figure}[t]
  \center
	\includegraphics[width=.70\textwidth]{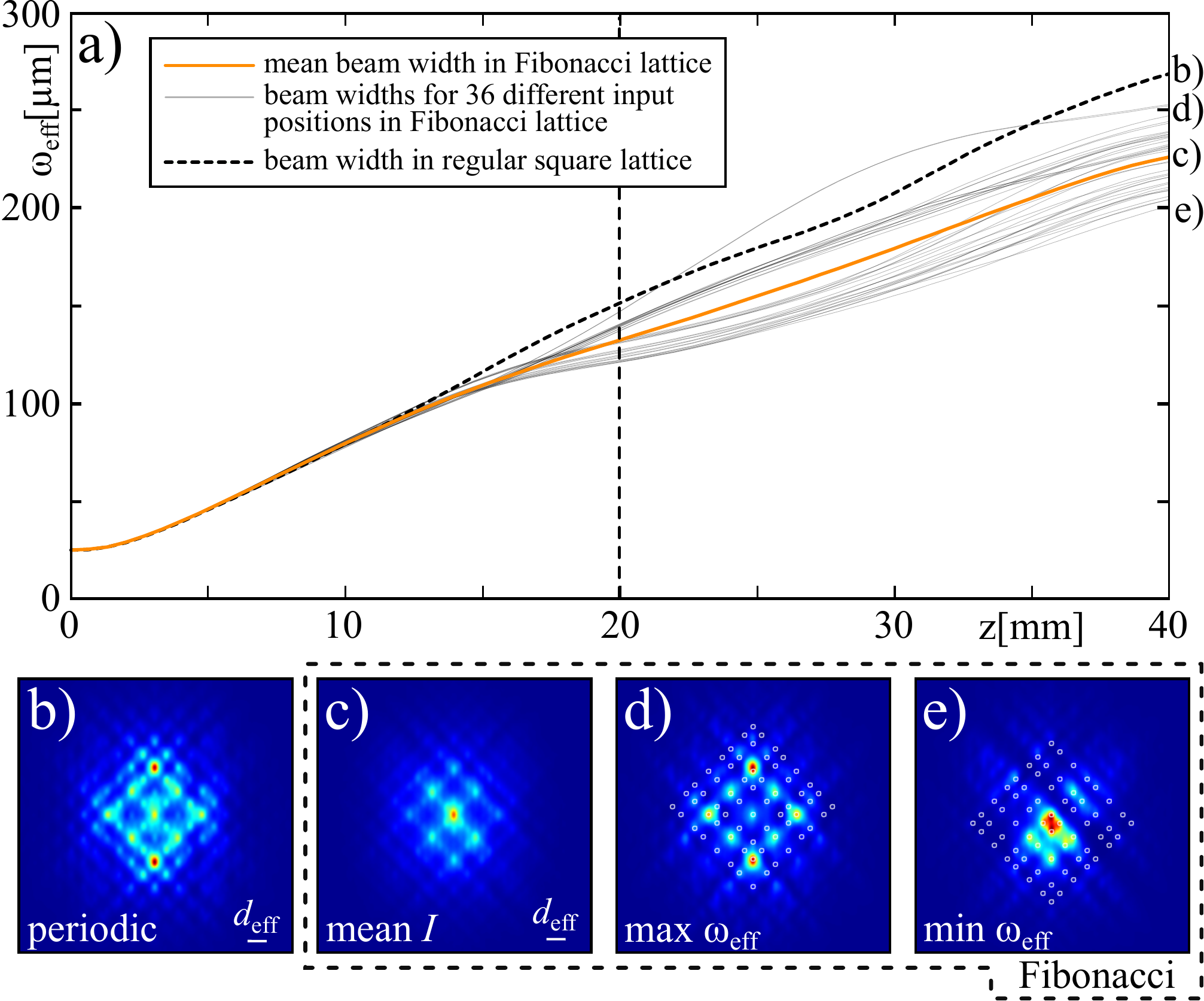}
	\caption{Comparison between strength of localization in periodic and Fibonacci lattice. (a)~Beam widths {$\indur{\omega}{eff}$} versus propagation distance {$z$}, (orange) mean beam width and (faint gray) single events in Fibonacci lattice; (dashed black) beam width in periodic lattice. Dashed vertical line at $z = \SI{20}{\milli\meter}$ indicates experimental propagation distance. Intensity distributions at $z = \SI{40}{\milli\meter}$ in (b)~periodic lattice, (c)~averaged intensity, (d)~maximum, and (e)~minimum effective beam width in Fibonacci lattice, each for $w_0 = \SI{14}{\micro\meter}$.
	}
	\label{fig:comparePeriodFib}
\end{figure}

Figure \ref{fig:comparePeriodFib} compares our numerical results of beam propagation in periodic lattices and aperiodic Fibonacci photonic lattices. 
In \fig{fig:comparePeriodFib}(a), averaged effective beam widths $\indur{\omega}{eff}$ along the propagation distance $z$ for Fibonacci and periodic lattice are presented for $w_0 = \SI{14}{\micro\meter}$, which corresponds to $\indur{\omega}{eff}(z = 0) \approx \SI{25}{\micro\meter}$.
The underlying numerical calculations consider a $15\times15$ lattice site arrangement in order to avoid surface influences.
The results in \fig{fig:comparePeriodFib}(a) indicate that $\indur{\omega}{eff}$ broadens slower during propagation in Fibonacci lattices than under periodic conditions.
While all effective beam widths are almost equal for the first $\SI{10}{\milli\meter}$, they separate for longer propagation distances since light starts to couple to next-nearest neighbors with respect to the input waveguide.
At this point, aperiodicity comes into play at the earliest.

Figure \ref{fig:comparePeriodFib}(b) presents the output after $\SI{40}{\milli\meter}$ of propagation in a periodic lattice.
A typical signature of discrete diffraction is prominent here.
The average of output intensity distributions for 36 different incident positions in Fibonacci lattice is presented in \fig{fig:comparePeriodFib}(c).
Compared with the periodic case, localization in Fibonacci lattices is evident, as highest intensity is predominantly located around the input center.
Such localization is weaker than expected in random media \cite{Boguslawski_AL} and diffraction is essentially suppressed in comparison with periodic lattices.

Again, the remarkably heterogeneous behavior for different input positions in the Fibonacci lattice can be found in \fig{fig:comparePeriodFib}(a) where particular arrangements promote beam widths that are even larger than for the periodic case, especially around $z = \SI{30}{\milli\meter}$.
In contrast, configurations holding effective beam widths considerably below the mean value naturally result from most localized modes.
To underline the diversity of localized modes occurring,  we depict in Figures \ref{fig:comparePeriodFib}(d,e) conditions of maximal and minimal $\indur{\omega}{eff}$, together confirming the relevance of the local waveguide arrangement conditions on light transport.

Obviously, transverse light transport is slowed down in average due to aperiodic conditions accounting for localized modes whose appearance indirectly indicate the presence of band gaps~\cite{DalNegro3}. 
Periodic structures, in contrast, bring forward extended propagation modes as more light is carried away from the input position due to discrete diffraction.

\section{Conclusions}
To conclude, we have observed light localization in optically induced two-dimensional Fibonacci photonic lattices that arises from the aperiodic property of the waveguide arrangement.
By experimentally and numerically analyzing linear propagation characteristics for various incident positions, we observed enhanced localization in Fibonacci photonic lattices compared with discrete-diffraction driven light transport in equivalent periodic realizations.
We identify this localization with the occurrence of highly localized modes due to aperiodic structural conditions with diminished order, indirectly indicating the existence of band gaps.

A very good agreement between experimental and numerical results allowed to additionally analyze the development during propagation of a wave package sent to corresponding lattices.
These numerical results consolidated our experimental observations and, moreover, gave deeper insights into propagation dynamics.
We further found that selecting the input position in such a deterministic aperiodic structure is highly crucial to the beam broadening during propagation.
Both, broader and smaller effective beam widths could be found among different scenarios compared to propagation in periodic waveguide arrangements.
Evidently, the shape of excited localized modes and thus the strength of localization are local rather than global properties in these kind of structures that have been subject of our investigations.

We are convinced that our results can be generalized to other kinds of aperiodic refractive index lattices, using the presented ideas and methods.

\section*{Acknowledgments}
This work is supported by the German Academic Exchange Service (Project 56267010) and Ministry of Education, Science and Technological Development, Republic of Serbia (Project OI 171036).

\end{document}